\documentclass[12pt]{iopart}
\usepackage{setstack}
\usepackage{graphicx}
\usepackage{dsfont}
\usepackage{mathptmx}

\newcommand{\PPR}[0]{P^{\mathrm{PR}}}
\newcommand{\PPRanti}{P^{\overline{\mathrm{PR}}}}
\newcommand{\PLL}[0]{P^{\mathrm{L}}}

\newcommand{\ket}[1]{\mbox{$ | #1 \rangle $}}
\newcommand{\bra}[1]{\mbox{$ \langle #1 | $}}

\begin{document}
\title{Couplers for non-locality swapping}
\author{Paul Skrzypczyk and Nicolas Brunner}
\address{H.H. Wills Physics Laboratory, University of Bristol, Tyndall Avenue, Bristol, BS8 1TL, United Kingdom}
\ead{\mailto{paul.skrzypczyk@bristol.ac.uk}, \mailto{n.brunner@bristol.ac.uk}}

\begin{abstract}
Studying generalized non-signalling theories brings insight to the foundations of quantum mechanics. Here we focus on a dynamical process in such general theories, namely non-locality swapping, the analogue of quantum entanglement swapping. In order to implement such a protocol, one needs to define a \emph{coupler}, which performs the equivalent of quantum joint measurements on generalized `box-like' states. Establishing a connection to Bell inequalities, we define consistent couplers for theories containing an arbitrary amount of non-locality, which leads us to introduce the concepts of perfect and minimal couplers. Remarkably, Tsirelson's bound for quantum non-locality naturally appears in our study.
\end{abstract}
\maketitle

\section{Introduction}

Quantum Mechanics (QM) is a non-local theory \cite{Bell64}, however not a maximally non-local one according to relativity \cite{PR}. More precisely there exist theories, containing more non-locality than QM, that still respect the no-signaling principle \cite{barrett,john}. The study of such theories has already provided a deeper understanding of the foundations of QM \cite{john,NS,barnum,prsinglet}, but one of the great remaining challenges is to find what physical principle, yet still unknown despite intensive research, limits quantum non-locality.

Interestingly, studying the communication properties of these models has already brought insight to this question. Notably, van Dam \cite{vanDam} showed that the paradigmatic example of such a generalized theory, characterized by maximally non-local correlations known as Popescu-Rohrlich (PR) boxes \cite{PR}, appears very unlikely to exist since it allows for a dramatic increase of communication power compared to QM. Essentially, in this model, all communication complexity problems become trivial. Indeed this is not the case in QM, and more generally it is strongly believed that communication complexity is not trivial in nature \cite{brassardNP,sanduNP}. This result was subsequently extended by Brassard et al.\ \cite{brassard} to a class of noisy PR boxes (isotropic PR boxes), and more recently it was shown that there exist correlations arbitrarily close to the set of classical correlations that also collapse communication complexity \cite{Distillation}. Moreover, Linden et al.\ \cite{noah} showed that isotropic PR boxes allow for non-local computation, a task for which quantum correlations offer no advantage compared to classical correlations. The remarkable fact about this last work is that it indicates a tight separation between quantum and post-quantum correlations.

Recently it was also suggested that the bound on quantum non-locality may be a consequence of the rich dynamics featured in QM. In particular, Barrett \cite{john} and Short et al. \cite{short} showed that a theory restricted only by the no-signaling principle allows only for poor (or classical) dynamics, suggesting ``\textit{a trade-off between states and measurements}" \cite{tonijohn}. QM might then appear to be the perfect compromise, allowing for both non-locality and rich dynamics.

In a recent paper \cite{emergence}, we (together with a third author) introduced the concept of a \emph{genuine} box, a particular way of restricting the set of allowed boxes in a generalized non-signalling theory. Inspired from the black-box approach to quantum correlations \cite{DevIndep}, we argued that the set of boxes to be considered for dynamical processes (such as joint measurements) must be restricted to genuine boxes. The idea is that these genuine boxes are the elementary states of the theory; all other boxes can then be constructed by adding classical circuitry. Remarkably, this restriction allows one to reintroduce quantum-like dynamics into the model, even in theories allowing for maximal non-locality.

The theory features then a new element, the \emph{coupler}, a device performing the analogue of a quantum joint measurement \cite{short}. The coupler enables non-locality swapping of PR boxes and teleportation. Astonishingly, quantum correlations, in particular Tsirelson's bound \cite{tsirelson2} of quantum non-locality, naturally emerged from the coupler \cite{emergence}.

In the present paper, we generalize the coupler of Ref. \cite{emergence} to theories allowing for limited non-locality. We also investigate the possibility  of varying the set of genuine boxes. For each case we derive couplers for non-locality swapping and study their properties. Remarkably, quantum correlations, more precisely Tsirelson's bound, will appear again in our study.

The paper is organized as follows. After reviewing general properties of couplers in \Sref{s:properties}, we indicate a strong analogy between couplers and Bell inequalities in \Sref{s:bell}. Taking advantage of this connection, we then extend the coupler of \cite{emergence} to theories with bounded non-locality, introducing perfect and minimal couplers in \Sref{s:gen}. Indeed an example of particular interest will be a theory where non-locality is bounded by Tsirelson's bound, similarly to quantum mechanics, presented in \Sref{s:quant}. Finally we highlight two occurrences where Tsirelson's bound naturally appears in \Sref{s:tsi}.

\section{Properties of couplers}\label{s:properties}

\begin{figure}[b]
   \begin{center}\includegraphics[scale=0.4]{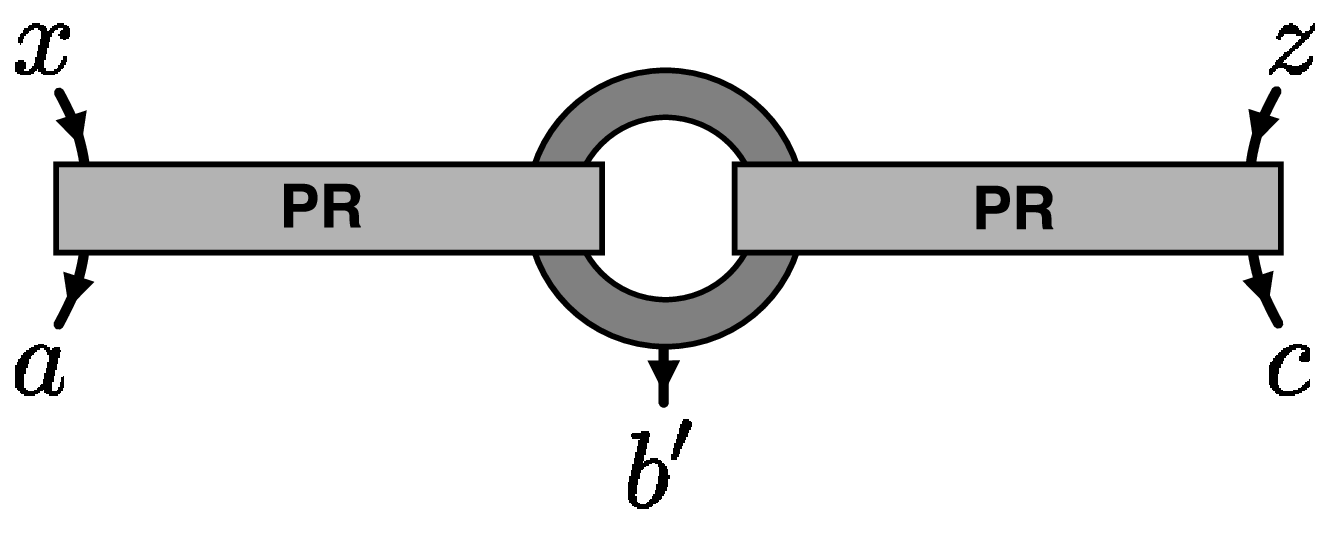}\end{center}
   \caption{Coupler for non-locality swapping. The coupler is the analogue of a quantum joint measurement. When applied to two PR boxes shared by Alice-Bob and Bob-Charlie, it enables non-locality swapping. More precisely, when the protocol succeeds ($b'=0$) the final box shared by Alice and Charlie $P(ac|xz)$ is non-local since it violates the CH inequality.}\label{swap}
 \end{figure}

Here we work in generalized non-signaling theories, where states are bipartite `box-like' states. In general boxes take inputs, $x,y\in \{0,1,...,m-1\}$, and provide outputs, $a,b\in \{0,1,...,n-1\}$. Each box is then characterized by a set a joint probabilities $P(ab|xy)$. To be valid the probability distribution must be non-signalling. A box is then either local, when its probability distribution can be reproduced by classical means only (shared randomness), or non-local when it violates a Bell inequality.

Let us now consider the scenario of non-locality swapping, the analogue of quantum entanglement swapping \cite{EntSwapping}. An observer, Bob, shares non-local boxes with both Alice and Charlie. The goal of the protocol is for Bob to establish non-local correlations between (initially uncorrelated) Alice and Charlie. In order to do this, Bob applies the coupler to his two boxes (see figure 1), which is the analogue of a quantum joint measurement \cite{short,john}. The coupler $\chi$  encompasses the inputs and outputs of his two boxes, and returns a single bit $b'$ \footnote{Note that more generally the coupler could feature more outputs; for instance two bits, like in the case of a quantum Bell state measurement. Here however, we will focus on the simplest case.}, thus implementing the following transformation:
\begin{equation}
    P(ab_1|xy_1)P(b_2c|y_2z) \stackrel{\chi}{\rightarrow} P(ab^\prime c|xz) \,.
\end{equation} where $P(ab_1|xy_1)$ is the initial box shared by Alice and Bob, $P(b_2c|y_2z)$ the initial box shared by Bob and Charlie. The final box shared by Alice and Charlie, given that the coupler returns $b'$, is $P(ac|xz b^\prime )$.

In order to be valid, the coupler must fulfill a certain number of requirements; we refer the reader to \cite{short} for more details on properties of couplers. First of all, the coupler must be non-signaling, in the sense that Bob cannot signal to Alice and Charlie by applying or not applying the coupler. Therefore, one must have
\begin{eqnarray}
    P(ac|xz) = \sum_{b^\prime}P(ab^\prime c|xz)  = \sum_{b'} P(b')P(ac|xzb') = P(a|x)P(c|z)\, .
\end{eqnarray}
Moreover, the coupler must be consistent when applied directly to any bipartite box allowed in the model. That is, one should check that the probability that the coupler outputs $b'$,
\begin{eqnarray}\label{toni}	P(b^\prime) = \sum_{b_1b_2y_1y_2} \chi(b^\prime, b_1b_2y_1y_2)P(b_1b_2|y_1y_2) \end{eqnarray} is a valid probability, i.e. $0 \leq P(b')\leq 1$. Note that since the coupler's action on a box is linear \cite{short}, it is sufficient to check this for  extremal allowed boxes only.

\section{Connection between couplers and Bell inequalities}\label{s:bell}

As just mentioned, the probability that the coupler outputs $b^\prime$ is a linear function of the box the coupler is applied to. Therefore it is convenient to rewrite \eref{toni} in vectorial form:
\begin{equation}
	P(b^\prime=0)  =\vec{\chi}\cdot\vec{P}
\end{equation}
Subsequently, the requirement that the coupler $\chi$ outputs with a valid probability when applied on the box $\vec{P}$ is given by
\begin{equation}\label{e:chidotp}
	0 \leq \vec{\chi}\cdot\vec{P} \leq 1
\end{equation} for all allowed boxes $\vec{P}$.

In \cite{short} \Eref{e:chidotp} was (rightly) interpreted as defining a polytope of couplers: the \emph{coupler polytope}. Its facets are given by the set of vectors $\vec{P}$; any consistent coupler is therefore a vector $\vec{\chi}$ inside the polytope. There is however a second possible interpretation of \eref{e:chidotp}, which consists in inverting the roles of the the vectors $\vec{P}$ and the vectors $\vec{\chi}$, thus defining another polytope, the dual of the coupler-polytope. This new polytope is in fact much more familiar; it is the (well-known) polytope of states \cite{pito}. In this representation, the vertices of the polytope are specified by the extremal boxes (vectors $\vec{P}$) while its facets are now associated to the couplers $\vec{\chi}$. The advantage of this second interpretation is that facets of the state polytope are, at least in some cases, well understood \cite{barrett}. Let us illustrate this connection by presenting two examples.

First, let us consider the set of all boxes consistent with the no-signaling principle. This set of boxes is known to form a polytope, the non-signalling polytope. The facets of this polytope are the \emph{positivity} facets (also called \emph{trivial} facets), which ensure that probabilities are positive. According to our interpretation of equation \eref{e:chidotp}, the extremal couplers are now associated to positivity facets. For the case of binary inputs and outputs, it has been shown that any valid coupler is a \emph{wiring}, that is a measurement which admits a classical description in terms  of inputting into each box and applying circuitry. Indeed wirings are not joint measurements. In Appendix A, we show how the wirings originate from the positivity facets. More generally, associating positivity facets to wirings strongly suggests that there are no joint measurements in any theory (arbitrary number of inputs and outputs) constrained only by the no-signaling principle, since in this case the state polytope has only positivity facets. This turns out to be correct as shown recently in \cite{tonijohn}

Second, let us consider the \emph{genuine} box model of Ref. \cite{emergence}. In this model, defined for boxes with binary inputs and outputs, the valid boxes form a restricted subset of the full non-signaling polytope. More precisely, the set of valid boxes consists of all local deterministic boxes, and a single (genuine) PR box (see below). In this case, the facets of the state polytope consist of the positivity facets, plus some Bell inequalities. Here it should be reminded that Bell inequalities correspond to the (non-trivial) facets of the state polytope when only local boxes are considered. Therefore in the genuine box model, by removing all but one of the PR boxes, it happens that all but one of the Bell inequalities reappear as facets of the state polytope. Specifically, we have the Clauser-Horne (CH) Bell-inequality \cite{CH}, given in the form of a scalar product as
\begin{equation}\label{CH}\nonumber
	\vec{\mathrm{CH}}\cdot\vec{P}(ab|xy) = P(11|00) + P(00|10) + P(00|01) - P(00|11)\,,
\end{equation}
with all local boxes satisfying $0\leq \mathrm{CH}\leq 1$. Notice that $0\leq \vec{\mathrm{CH}}\cdot\vec{P}(ab|xy)$ is now a facet of the polytope of genuine boxes. As expected, there is a new measurement corresponding to this CH Bell inequality facet. This measurement is the coupler presented in Ref \cite{emergence}, which implements the analogue of a quantum joint measurement, and enables non-locality swapping of PR boxes. The action of this coupler on any allowed box $P(ab|xy)$ was found to be
\begin{equation}\label{maincoupler}
    P(b^\prime=0) = \vec{\chi}\cdot \vec{P}(ab|xy) = \case{2}{3}\vec{\mathrm{CH}}\cdot\vec{P}(ab|xy) \,.
\end{equation}
The proportionality factor can be easily understood: since $0\leq \mathrm{CH}\leq \frac{3}{2}$ for any allowed box in the genuine box model (the PR box having a CH value of $\frac{3}{2}$), the constant ensures that $0 \leq \vec{\chi}\cdot \vec{P}(ab|xy)\leq 1$ as desired. Thus the measurement which implements non-locality swapping corresponds to a Bell inequality. Intuitively, this should be understood in the following way: a joint measurement is a global action, so it is natural to associate it with a Bell inequality, since the latter reveals a joint property of a box, namely the amount of non-locality it contains. Finally, note that the other Bell inequality facets (symmetries of the CH inequality) that reappear in the genuine box model cannot be associated to consistent measurements, since they allow one to create disallowed boxes.

To summarize we have seen two examples illustrating the close relation existing between couplers and facets of the state polytope. On the one hand when we consider all boxes consistent with the no-signaling principle (i.e. the full non-signalling polytope), all facets of the state polytope are positivity facets, which generate measurements with a classical description, i.e. wirings. On the other hand, in the genuine box model, where the polytope has exposed CH Bell inequality facets, an inherently joint measurement emerges: the coupler for non-locality swapping. Moreover the coupler corresponds to one of the exposed CH Bell inequality facets.

In general, we may ask what happens for other models, for instance if we take a set of boxes which is not the full non-signaling polytope yet contains no exposed CH Bell inequality facets. In such a setting it is found that the the (non-trivial) facets of the polytope are of a mixed type, that is a mixture of positivity and Bell inequality facets. Therefore there exist measurements which cannot be understood as classical wirings. Whether or not such `noisy' Bell inequality facets are also useful for non-locality swapping will be the focus of the rest of this paper.

\section{Generalized couplers}\label{s:gen}

The previous connection can now be built upon substantially to look at the task of non-locality swapping in a more general setting. In \cite{emergence} we restricted the set of genuine boxes to the (local) deterministic boxes
\begin{equation}\label{local}
\PLL_{\alpha \beta \gamma \delta}(ab|xy) =
\cases{
1 & $ a = \alpha x \oplus \beta$ \,\, , \,\, $b = \gamma y \oplus \delta $ \\
0 & otherwise
}
\end{equation} parameterized by $\alpha,\beta,\gamma,\delta \in \{0,1\}$, and added a single non-local vertex, the PR box:

\begin{equation}\label{PR}
\PPR (ab|xy) =
\cases{
\case{1}{2} & $ a \oplus b = xy$ \\
0 & otherwise
}
\end{equation} where $\oplus$ denotes addition modulo 2.

\begin{figure}
  \begin{center}\includegraphics[scale=0.4]{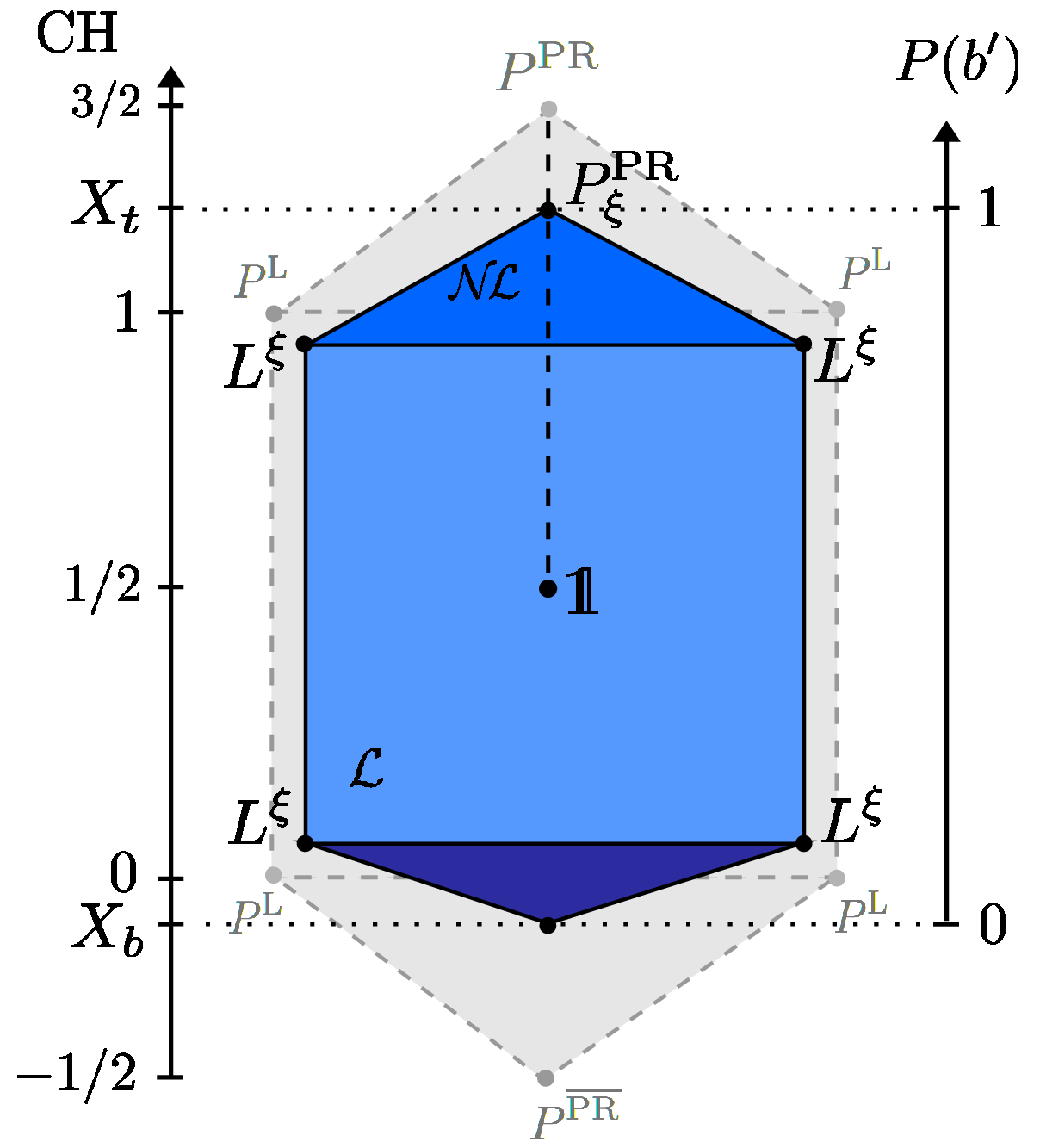} \end{center}
  \caption{The set of boxes allowed in our model. Here we study theories with limited non-locality. More precisely, non-locality is upper bounded by $X_{\mathrm{t}}$ and lower bounded by $X_{\mathrm{b}}$; for perfect couplers, the genuine boxes must be taken to be $L^{\xi}$ and $\PPR_{\xi}$ (see text). In order to output with a valid probability when applied on any allowed box, the coupler (right scale) must be a re-scaling of the CH inequality (left scale). Note that the asymmetry of the restricted polytope ($X_{\mathrm{b}}<1-X_{\mathrm{t}}$) will turn out to be a necessary condition for the existence of couplers.}\label{polytope}
\end{figure}

A natural extension of this previous analysis consists of varying the set of genuine boxes. There are two interesting directions to investigate (see figure \ref{polytope}):
\begin{enumerate}
 \item Limit the amount of non-locality allowed in the theory; that means that the genuine PR box \eref{PR} may be replaced by a noisy box.
\item One may add another genuine non-local box, violating the lower bound of the CH inequality. This will bring back into the study non-local boxes sitting in the lower  region  of the polytope, which was completely discarded in our previous study \cite{emergence}.
\end{enumerate}

Here we shall choose all non-local genuine boxes to be isotropic, though a similar study may be done for more general genuine boxes. Isotropic boxes have the form
\begin{eqnarray}\label{isotropic} \PPR_{\xi} = \xi \PPR + (1-\xi) \PPRanti \end{eqnarray} where $\PPRanti$ is the `anti-PR' box:
\begin{equation}\label{antiPR}
\PPRanti (ab|xy) =
\cases{
\case{1}{2} & $a \oplus b \oplus 1= xy$ \\
0 & otherwise
}
\end{equation}
One has $\vec{\mathrm{CH}}\cdot\vec{\PPRanti}=-\frac{1}{2}$, and $\vec{\mathrm{CH}}\cdot\vec{\PPR_{\xi}}=2\xi -\frac{1}{2}$.

Here we will choose the set of genuine boxes by fixing the amount of non-locality allowed in the theory. More precisely we will define an upper bound $X_{\mathrm{t}}$ as well as a lower bound $X_{\mathrm{b}}$ on the CH value of authorized boxes (see figure 2).

To associate a coupler to a given choice of $X_{\mathrm{t}}$ and $X_{\mathrm{b}}$ we recall that our previous coupler \cite{emergence} returned $b^\prime=0$ with a probability proportional to the CH value of the box it is applied to (see \eref{maincoupler}). For the PR box -- the box leading to the largest CH value -- the output $b^\prime = 0$ is deterministically returned, while $b^\prime = 1$ is always returned for boxes sitting on the lower CH facet -- boxes with the smallest CH value.

This suggests the following generalization. The coupler should return the output $b^\prime = 0$ deterministically for boxes with $\mathrm{CH} = X_{\mathrm{t}}$ and $b^\prime = 1$ for boxes with $\mathrm{CH} = X_{\mathrm{b}}$ (see figure 2). In practice this means that the coupler will not simply be proportional to the CH value, but given by the linear function
\begin{equation}\label{coupler}
	\vec{\chi} = \frac{1}{X_{\mathrm{t}}-X_{\mathrm{b}}}\left(\vec{\mathrm{CH}} - X_{\mathrm{b}}\vec{\chi}_D\right)
\end{equation}
where $\vec{\chi}_D$ is the \emph{deterministic} coupler which always outputs $b^\prime = 0$ and acts as an identity element \footnote{One possibility for the deterministic coupler is: $\vec{\chi}_D \cdot \vec{P} = P(00|00)+P(01|00)+P(10|00)+P(11|00)=1$ for all normalized $\vec{P}$.}. Note also that \eref{coupler} is a shift of $-X_{\mathrm{b}}$ of the CH value, followed by a re-scaling by a factor of $1/(X_{\mathrm{t}}-X_{\mathrm{b}})$.

Let us stress that these couplers are probabilistic, similarly to quantum partial Bell state measurements. We will refer to $P(b^\prime=0)$ as the \emph{success probability} of the coupler as when Bob obtains the outcome $b^\prime = 0$ the desired (non-local) box is created between Alice and Charlie. Note that in case the coupler fails ($b'=1$), Alice and Charlie are left with a local box\footnote{We show in Appendix B that it is always the case that when the coupler outputs $b' = 1$ that the box is necessarily local.}. It can be shown that the success probability of the coupler \eref{coupler}, when Bob shares two initial $\PPR_{\xi}$ boxes with Alice and Charlie, is given by
\begin{eqnarray}\label{success} P(b'=0)= \frac{1-2X_{\mathrm{b}}}{2(X_{\mathrm{t}}-X_{\mathrm{b}})} \end{eqnarray}
 which notably is independent of $\xi$.
In the case of success, the CH value of the final box shared by Alice and Charlie is
\begin{eqnarray}\label{CH_AC} \vec{\mathrm{CH}} \cdot \vec{P}(ac|xz) = \frac{1}{1-2X_{\mathrm{b}}}\left( 2\xi-1  
\right)^2 + \frac{1}{2} \end{eqnarray} These two last relations can now be used to characterize the entire class of couplers -- details of their derivations can be found at the end of the paper in Appendix B.

Consistency requires that the final box shared between Alice and Charlie should not be more non-local than the original boxes shared between Alice-Bob and Bob-Charlie, since this would enable the creation of non-locality. On the other hand, for non-locality to be swapped we also require the final box to be non-local. In the next subsections, we show that valid couplers, i.e. satisfying these requirements, are characterized by the relations:

\begin{eqnarray}\label{main} \frac{1}{2} -(X_{\mathrm{t}}-\frac{1}{2})^2   < X_{\mathrm{b}} \leq \frac{\frac{3}{2}-X_{\mathrm{t}}}{2}\,. \end{eqnarray}

\subsection{Perfect couplers}

An important requirement for the coupler is that it does not allow one to create non-locality, otherwise the study of models with restricted non-locality would be pointless. Mathematically, this translates to the condition  \begin{eqnarray} \vec{\mathrm{CH}} \cdot \vec{P}(ac|xz) \leq X_{\mathrm{t}}\,. \end{eqnarray} Inserting $2\xi-1=X_{\mathrm{t}}-\frac{1}{2}$ into \eref{CH_AC}, this leads to \begin{eqnarray}\label{perfect} X_{\mathrm{b}} \leq \frac{\frac{3}{2}-X_{\mathrm{t}}}{2}  \end{eqnarray} the right hand inequality of \eref{main}.

We call a coupler reaching the upper bound of inequality \eref{perfect} a \emph{perfect} coupler. When Bob applies such a perfect coupler (and the swapping succeeds), the final box of Alice and Charlie is as non-local as the two initial boxes shared by Alice-Bob and Bob-Charlie; starting from two $\PPR_{\xi}$ boxes, Alice and Charlie get a $\PPR_{\xi}$, where $\PPR_{\xi}$ is the most non-local box allowed in the model, i.e. $X_{\mathrm{t}}=2\xi-\frac{1}{2}$. The coupler presented in Ref. \cite{emergence} (given by $X_{\mathrm{b}}=0$, $X_{\mathrm{t}}=\frac{3}{2}$) is a perfect coupler -- it swaps two PR boxes to a PR box -- and indeed saturates inequality \eref{perfect}. For perfect couplers, the probability of success, i.e. of obtaining the outcome $b'=0$, turns out to be equal to $\frac{1}{3}$, independently of $X_{\mathrm{t}}$.

Remarkably, a perfect coupler can be found for any model with limited non-locality (see figure 3). However the existence of such a perfect coupler imposes restrictions on the set of genuine local boxes, since $X_{\mathrm{b}}>0$ when $1<X_{\mathrm{t}}<\frac{3}{2}$. This means that the deterministic boxes sitting on the lower CH facet are no longer authorized. One could then argue that, since any local box can be simulated by Alice and Bob from shared randomness, a perfect coupler can never be consistently defined for theories with limited non-locality (i.e with $1<X_{\mathrm{t}}<\frac{3}{2}$).

However, it should be pointed out that boxes (local or non-local) are resources, and that being able to simulate a box is not equivalent to actually holding the box, much in the same way that being able to simulate a quantum state is not equivalent to actually holding the state. Indeed this observation is particularly important when considering dynamical processes, such as joint measurements. From this point of view it is crucial to distinguish the set of boxes that Alice and Bob can actually prepare in a theory, from those they can only simulate. Importantly, while it is necessary to require consistency of the coupler when applied onto the first ones, it is not necessary to ask for consistency for the second ones.

In particular considering models with limited non-locality, we shall see below that only noisy local boxes can actually be created (starting from a noisy non-local PR box). Therefore noiseless deterministic boxes do not have to be considered as genuine, and it is thus not necessary to require that the coupler acts consistently on them.

\subsubsection{Noisy local boxes}\label{s:localnoisy}

Let us think about how single-party boxes can be obtained starting from an initial bipartite non-local box, in the most restricted scenario, where Alice and Bob are only allowed to input in their boxes and then obtain an output.

For clarity, we start with the case of a maximally non-local theory, i.e. the PR box. Here Alice and Bob can create (noiseless) deterministic boxes in the following way. Suppose Alice and Bob share a PR box, and Alice, after inputting $x$ into the box and getting output $a$, sends $x$ and $a$ to Bob. Then Bob holds the deterministic (single-party) box $\PLL_{xa}$ (i.e. $b = xy \oplus a$). This explains why all deterministic local boxes must be considered as genuine (in addition to the PR box) in the genuine box model of Ref. \cite{emergence}.

Now, in a theory with limited non-locality, i.e. where extremal non-local boxes are noisy PR boxes $\PPR_{\xi}$, the local boxes obtained by such a procedure are not deterministic but noisy (see figure 2). These boxes are given by \begin{eqnarray} L_{\alpha \beta}^{\xi}= \xi \PLL_{\alpha\beta} + (1-\xi) \PLL_{\alpha,\beta\oplus1} \end{eqnarray} One can show that any bipartite local boxes obtained from these noisy local boxes satisfy
\begin{eqnarray} Z_{\mathrm{b}} \leq \vec{\mathrm{CH}} \cdot \vec{L}_{\alpha\beta\gamma\delta}^{\xi} \leq Z_{\mathrm{t}} \end{eqnarray} where $Z_{\mathrm{b}} = \frac{1}{2} \left( 1-(2\xi-1)^2 \right)$, $Z_{\mathrm{t}}= 1- Z_{\mathrm{b}}$, and $L_{\alpha\beta\gamma\delta}^{\xi} \equiv L_{\alpha\beta}^{\xi} L_{\gamma\delta}^{\xi}$

\begin{figure}
  \begin{center}\includegraphics[scale=0.45]{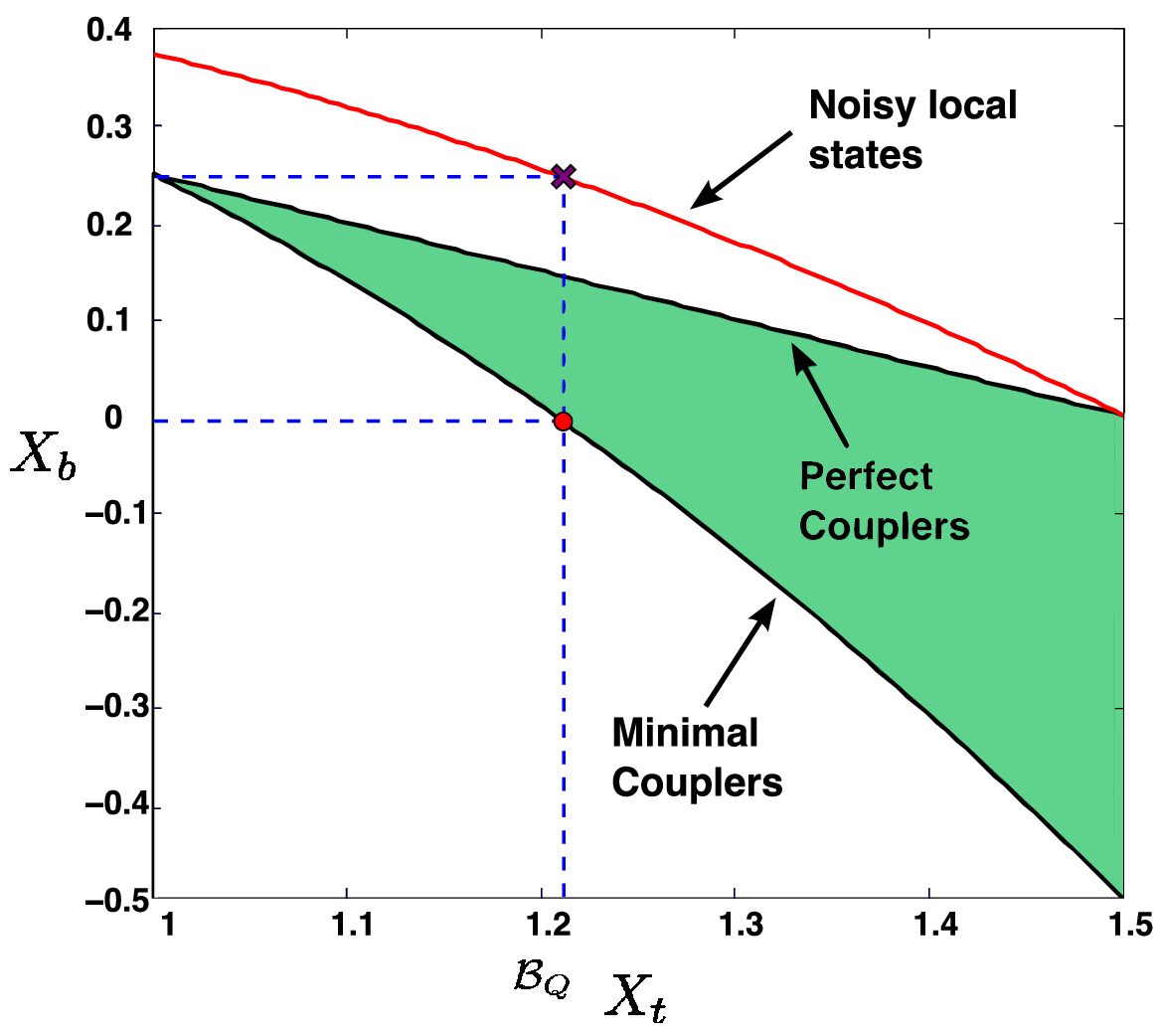}\end{center}
  \caption{Couplers (defined by $X_{\mathrm{b}}$ and $X_{\mathrm{t}}$) for non-locality swapping in theories where non-locality is bounded; more precisely the CH value of any allowed box is upper bounded by $X_{\mathrm{t}}$. The shaded region, delimited by perfect and minimal couplers, represents all valid couplers. Note that a perfect coupler can be validly defined for all theories, since the noisy local boxes (see text) are allowed (their CH value is always larger than $X_{\mathrm{b}}$). The minimal coupler allows one to keep all deterministic boxes if and only if the theory is post-quantum (dot). Furthermore, in a theory with an arbitrarily small amount of non-locality, the noisy boxes are obtained from a Tsirelson box (cross), the maximally non-local box allowed by QM.}\label{graph}
\end{figure}

Importantly, for a perfect coupler one has that $X_{\mathrm{b}}\leq Z_{\mathrm{b}}$ (indeed $Z_{\mathrm{t}}<X_{\mathrm{t}}$), thus ensuring that the coupler acts consistently on any local boxes that can be created in the model. Therefore the perfect coupler can be consistently defined in any model with limited non-locality, i.e. with $1<X_{\mathrm{t}}\leq \frac{3}{2}$. Note that local boxes sitting on the upper CH facet must also be made noisy, otherwise the coupler runs into inconsistencies.

Let us point out that, except for the case of a fully non-local theory, the CH value of the noisy local boxes do not reach the minimal authorized CH value $X_{\mathrm{b}}$ (see figure 3). More precisely, one has that $X_{\mathrm{b}} < Z_{\mathrm{b}}$ when $1<X_{\mathrm{t}}< \frac{3}{2}$. Indeed one could have expected to find that these two values would in fact coincide (i.e. $X_{\mathrm{b}}=Z_{\mathrm{b}}$ for all $X_{\mathrm{t}}$), thus giving a natural motivation for restricting the set of genuine local boxes, but this is not the case. An interesting open question would be to explain this discrepancy.


\subsection{Minimal couplers}

A second requirement for the coupler is that it swaps non-locality, i.e. When Bob applied the coupler and is successful in swapping non-locality using two copies of the most non-local boxes allowed in the model ($\vec{\mathrm{CH}} \cdot \vec{\PPR_\xi}=X_{\mathrm{t}})$, non-locality -- even an arbitrarily small amount -- is swapped to Alice and Charlie. This implies

\begin{eqnarray}  \vec{\mathrm{CH}} \cdot \vec{P}(ac|xz) >1  \end{eqnarray} which leads to

\begin{eqnarray} X_{\mathrm{b}} > \frac{1}{2} - (X_{\mathrm{t}}-\frac{1}{2})^2  \end{eqnarray} the left hand inequality of \eref{main}.

Here $X_{\mathrm{b}}$ represents, roughly speaking, the tolerable amount of boxes sitting in the lower region of the polytope. In particular, deterministic boxes can be kept if and only if the model is post-quantum, i.e. $X_{\mathrm{t}}>B_Q$ where $B_Q=\frac{1}{2}+\frac{1}{\sqrt{2}}$ is Tsirelson's bound \footnote{Note that here we consider the CH inequality -- for the CHSH inequality, one has the well known value $B_Q=2\sqrt{2}$.}. Interestingly, it is precisely when the model becomes quantum that the deterministic boxes must be made noisy ($X_{\mathrm{t}}=B_Q$ implies $X_{\mathrm{b}}>0$). The probability of success for the minimal coupler is bounded by $P(b'=0)<(X_{\mathrm{t}}-1/2)/(X_{\mathrm{t}}+1/2)$; note that $P(b'=0) \rightarrow \frac{1}{2}$ when $X_{\mathrm{t}} \rightarrow \frac{3}{2}$.

Interestingly, the minimal coupler in a model restricted only by non-signaling (i.e. $X_{\mathrm{t}}=\frac{3}{2}$) is given by $X_{\mathrm{b}} >-\frac{1}{2}$. Thus non-locality swapping is possible as long as the anti-PR box is discarded. More generally, it can be seen from figure \ref{graph} that a theory with symmetric non-locality (that is with $X_{\mathrm{b}} = 1-X_{\mathrm{t}}$) does not allow the existence of couplers, therefore enforcing the idea that some boxes must be discarded in order to get interesting dynamics.

\section{Quantum Case}\label{s:quant}

 Of particular interest is a theory which features the same amount of non-locality as in quantum mechanics. Here non-locality is limited by Tsirelson's bound $X_{\mathrm{t}}=B_Q $. The perfect `quantum' coupler is given by $X_{\mathrm{b}} = \frac{1}{2}(1-\frac{1}{\sqrt{2}})$. It prevents non-locality swapping when the two initial isotropic boxes are such that $\vec{\mathrm{CH}} \cdot \vec{\PPR_\xi} \leq \frac{1}{2}+ 2^{-\frac{3}{4}} $. It is worth mentioning that in quantum mechanics, Werner states, $\rho_w = w \ket{\psi^-}\bra{\psi^-}+ (1-w) \frac{\mathds{1}}{4}$, cannot be swapped under a similar condition, namely $\vec{\mathrm{CH}} \cdot \vec{P}_{\rho_w} \leq \frac{1}{2}+ 2^{-\frac{3}{4}} $. Note however that the perfect quantum coupler has a success probability of $\frac{1}{3}$, whereas a quantum partial Bell state measurement (here basically the projection onto the antisymmetric subspace) succeeds with probability $\frac{1}{4}$. Finally, the minimal quantum coupler can be associated to the perfect coupler for a PR box (see below).

\section{Perfect vs minimal and Tsirelson's bound}\label{s:tsi}

In this section we discuss the relation between perfect and minimal couplers, and show that Tsirelson's bound for quantum non-locality naturally emerges from it on two occasions.

Let us first point out that perfect and minimal couplers are directly related. In a model with a given amount of non-locality $X_{\mathrm{t}}$, the perfect coupler is characterized by $X_{\mathrm{b}}= (3/2 - X_{\mathrm{t}})/2$. Because of the linearity of the coupler, all non-local boxes with $ \vec{\mathrm{CH}} \cdot \vec{P^{\mathrm{PR}}_{\xi}}<X_{\mathrm{t}}$ are swapped to a noisier box $P^{\mathrm{PR}}_{\xi'}$ with $\xi'<\xi$. At some point, the boxes become too noisy and forbid non-locality swapping with the perfect coupler: let us denote the box at the threshold $\PPR_{th}$. Then it follows that the coupler defined by $X_{\mathrm{t}}=\vec{\mathrm{CH}}\cdot \vec{\PPR_{th}}$ and $X_{\mathrm{b}}$ is a minimal coupler. In other words, the point where a perfect coupler stops to swap corresponds to a minimal coupler.

Astonishingly, this implies that the perfect coupler for a PR box (i.e. $X_{\mathrm{t}}=\frac{3}{2}$, $X_{\mathrm{b}}=0$) corresponds to the minimal quantum coupler (i.e. $X_{\mathrm{t}}=B_Q$, $X_{\mathrm{b}} \to 0$). This is a way of rephrasing the result of \cite{emergence}; though there the correspondence could be generalized to a whole section of the polytope. Let us stress that this link is remarkable, since it relates a dynamical process in a very natural generalized theory directly to quantum correlations.

Next let us point out another occurrence where Tsirelson's bound naturally appears in our study. In a theory containing a vanishing amount of non-locality ($X_{\mathrm{t}} \rightarrow 1$), the perfect and minimal couplers coincide, as can be seen from figure \ref{graph}. This is intuitive since the theory allows only for very weakly non-local boxes. In this regime, the coupler is characterized by $X_{\mathrm{b}} \rightarrow \frac{1}{4}$; therefore the noisy local boxes $L^{\xi}$ must satisfy the condition that $\vec{\mathrm{CH}} \cdot \vec{L}^{\xi}\geq \frac{1}{4}$. Now, a natural question is the following. What non-local box is required in order to obtain such noisy local boxes (i.e. satisfying $\vec{\mathrm{CH}} \cdot \vec{L}^{\xi} = \frac{1}{4}$) from the procedure described previously (in \Sref{s:localnoisy}). The answer is that this box must be the Tsirelson box, that is the isotropic PR box $\PPR_{\xi}$ satisfying $\vec{\mathrm{CH}}\cdot \vec{\PPR_\xi}=B_Q$.

This second link is astonishing since it involves not only the coupler, but also the procedure for creating noisy local boxes explained previously. In this sense it is also clearly different from the first connection we mentioned above.

Let us stress that, at the moment, both of these connections remain completely mysterious to us. Nevertheless we believe they might be related to some physical principle potentially restricting quantum non-locality.

\section{Conclusion}

In summary, we presented a study of generalized couplers for non-locality swapping. We started by pointing out a strong connection between couplers and Bell-type inequalities. This led us to associate (trivial) positivity facets with classical measurements, so-called wirings, and Bell inequalities with joint measurements. Then, taking advantage of this connection we presented a general class of couplers for theories with limited non-locality. This allowed us to introduce two important classes of couplers, namely perfect and minimal couplers. Finally we discussed the quantum case and presented two appearances of Tsirelson's bound in our study

To conclude, we would like to point out some interesting open questions. First concerning the connection between couplers and Bell inequalities. There exists in fact another type of inequality -- apart from trivial and Bell inequalities. These are Bell-type inequalities allowing the use of some non-local resource \cite{bacon,NJP,JMP}, such as classical communication or non-local boxes. Interesting couplers may also appear from such inequalities. Second it would be worth studying more general scenarios, especially those featuring more measurement inputs. As noted in \cite{short}, the case of three settings is of particular interest, since quantum tomography of qubits requires three measurements. Next, concerning couplers, it would be nice to find a coupler performing the analogue of a complete (quantum) Bell state measurement, where all eigenstates are entangled. Another point is to see whether the existence of couplers have implications for information theoretic tasks in generalized non-signalling theories, for instance for bit commitment \cite{commit_wolf,commit_buhrman,commit_barnum,commit_short}, or non-locality distillation \cite{Foster,Distillation,toni_pur}. Finally, the biggest question is definitely to find why quantum correlations and couplers seem to be so intimately related.

\section{Acknowledgements}
The authors are grateful to J.\ E. Allcock, J.\ Barrett, T.\ S.\ Cubitt, A.\ R.\ U.\ Devi, N.\ Gisin, W.\ Matthews, S.\ Popescu, V.\ Scarani, A.\ J.\ Short and J.\ Wullschleger for many insightful discussions. P. S. acknowledge support through the UK EPSRC project `QIP IRC'. N. B. acknowledges financial support by the Swiss
National Science Foundation (SNSF).

\appendix

\section{Construction of wirings from positivity facets}
Here we show explicitly the connection pointed out in \Sref{s:bell}, between the extremal measurements (valid for all non-signalling boxes) and the positivity facets of the full non-signalling polytope. We recall that measurements must satisfy the constraint
\begin{equation}\label{e:chidotp2}
	0 \leq \vec{\chi}\cdot\vec{P} \leq 1
\end{equation} for all non-signalling boxes $\vec{P}$. Here we focus on the case of binary inputs and outputs. The case of more inputs and/or outputs should be a straightforward generalization; note however that for more than two parties, the generalization does not hold since there exist in this case extremal measurements that are not wirings \cite{tonijohn}. It was previously shown in \cite{short} that there are 82 extremal consistent measurements, all of which are wirings. Below we show how to construct all of these wirings starting from the facets of the full non-signalling polytope.

Boxes are represented by 16-dimensional real vectors. The components of the vector $\vec{P}$ specifying a box are the 16 joint probabilities $P(ab|xy)$. The probabilities being subjected to linear constraints (normalisation and no-signalling) it turns out that the polytope of non-signalling boxes lives in an 8-dimensional hyperplane. One of these linearities ensures normalization:
\begin{equation}
	\sum_{a,b}P(ab|xy) = 1 \,.
\end{equation}
This linearity provides the first measurement, the (trivial) {\sc deterministic} wiring $\vec{\chi}_{\rm D}$, which satisfies $\vec{\chi}_{\rm D}\cdot \vec{P} =1$, for all $\vec{P}$; the outcome $b'=0$ is deterministically outputted for any normalised box. Thus, this wiring acts as the identity. Note that one also gets the opposite {\sc deterministic} wiring, which outputs $b'=1$ for all $\vec{P}$, which is the origin of the coupler polytope.

Now, the 16 positivity facets of the non-signalling polytope, which ensure that each joint probability is positive, are given by $P(ab|xy)\geq 0$. Indeed, one also has that $P(ab|xy)\leq 1$, by combination of the normalisation linearity and positivity facets. Therefore to each positivity facets corresponds a valid measurement; in fact each positivity facet translates into an {\sc and} wiring, $\chi_{AND}^j$ with $j\in \{1,...,16\}$. For instance the facet $P(11|00)\geq 0$ corresponds to the {\sc and} wiring characterized by first inputting $x=y=0$ into the box and then outputting $b'=0$ iff $a=b=1$, i.e. $b'=ab\oplus 1$.

Then, starting from the deterministic wiring $\chi_D$ and any $\chi_{AND}^j$, one can construct the following wiring: $\tilde{\chi}_{AND}^j \equiv \chi_{D}-\chi_{AND}^j$ with $j\in \{1,...,16\}$. This is clearly a valid measurement, since $0 \leq \tilde{\chi}_{AND}^j \leq 1$ for all boxes $\vec{P}$. Note that the wirings $\tilde{\chi}_{AND}$ correspond also to {\sc and} wirings, and are extremal; they are a non-convex combination of extremal wirings.

So far we obtained 34 wirings: 2 deterministic and 32 {\sc and}. Now, the 48 remaining wirings are constructed as follows. Notice that the non-convex combination $\chi_{jk} \equiv \chi_{AND}^j + \chi_{AND}^k$, with $j,k\in \{1,...,16\}$, is not a valid measurement iff there is a box $\vec{P}$ such that $\chi_{AND}^j \cdot \vec{P}=1$ and $\chi_{AND}^k \cdot \vec{P}=1$. Then it is straightforward to check that the remaining 48 wirings, of which there are 8 {\sc one-sided}, 8 {\sc xor} and 32 {\sc sequential}, are generated in this way. Again, the wirings $\chi_{jk}$ are extremal; they are non-convex combinations of the extremal wirings $\chi_j$ and $\chi_k$.

\section{Derivation of coupler action}
We recall that the action of a coupler is a linear transformation of the form
\begin{equation}
	P(ab_1|xy_1)P(b_2c|y_2z) \stackrel{\chi}{\rightarrow} P(ab^\prime c|xz) \,.
\end{equation}
We are interested in the specific case of applying the coupler \eref{coupler} to isotropic PR boxes \eref{isotropic}. First it is advantageous to realise that the probability of success $P(b^\prime = 0)$ is determined entirely by the box that Bob holds locally, i.e.
\begin{equation}
	P(b^\prime=0) = \vec{\chi}\cdot\vec{P}(b_1b_2|y_1y_2).
\end{equation}
Since isotropic PR boxes have uniform marginals, independently of $\xi$, Bob always holds the maximally mixed box $\mathds{1}(b_1b_2|y_1y_2)$ and therefore we find
\begin{eqnarray}
\fl	P(b^\prime = 0) = \vec{\chi}\cdot\mathds{1}(b_1b_2|y_1y_2)
	= \frac{1}{X_{\mathrm{t}}-X_{\mathrm{b}}}\left(\vec{\mathrm{CH}} - X_{\mathrm{b}}\vec{\chi}_D\right)\cdot\mathds{1}(b_1b_2|y_1y_2)
	= \frac{1-2X_{\mathrm{b}}}{2(X_{\mathrm{t}}-X_{\mathrm{b}})},
\end{eqnarray}
where we used the fact that $\vec{\mathrm{CH}}\cdot\mathds{1} = \frac{1}{2}$.

To find the final box prepared between Alice and Charlie upon Bob obtaining the outcome $b^\prime = 0$ (i.e. when the coupler succeeds in swapping non-locality) we must use the facts that (i) the coupler \eref{coupler} is a linear combination of two couplers that both act linearly (ii) the isotropic PR boxes \eref{isotropic} are convex combinations of the PR and anti-PR box. Therefore the only actions we need to know are the following
\begin{equation}
\eqalign{\PPR_{AB_1}\PPR_{B_2C} \stackrel{\vec{\chi_{\mathrm{CH}}}}{\rightarrow} \frac{1}{2}\PPR_{AC} \qquad \PPR_{AB_1}\PPRanti_{B_2C} \stackrel{\vec{\chi_{\mathrm{CH}}}}{\rightarrow} \frac{1}{2}\PPRanti_{AC}\cr
\PPRanti_{AB_1}\PPR_{B_2C} \stackrel{\vec{\chi_{\mathrm{CH}}}}{\rightarrow} \frac{1}{2}\PPRanti_{AC} \qquad 
\PPRanti_{AB_1}\PPRanti_{B_2C} \stackrel{\vec{\chi_{\mathrm{CH}}}}{\rightarrow} \frac{1}{2}\PPR_{AC}} \qquad \mathrm{when  }\quad b^\prime = 0
\end{equation}
Note that any combination of $\PPR$ and $\PPRanti$ is taken to the maximally mixed box when the deterministic coupler ${\vec{\chi}_D}$ is applied. Upon expanding all the terms and using relations (B.4) the final box, $P_{\mathrm{S}}(ac|xz)$, is found to be, when $b^\prime = 0$
\begin{eqnarray}
	\fl P_{\mathrm{S}}(ac|xz) = \frac{1}{1-2X_{\mathrm{b}}}\bigg\{[1-2\xi(1-\xi)-X_{\mathrm{b}}]\PPR(ac|xz) + [2\xi(1-\xi)-X_{\mathrm{b}}]\PPRanti(ac|xz)\bigg\}
\end{eqnarray}
which has CH value
\begin{eqnarray} 
	\vec{\mathrm{CH}} \cdot \vec{P}_{\mathrm{S}}(ac|xz) = \frac{1}{1-2X_{\mathrm{b}}}\left( 2\xi-1 \right)^2 + \frac{1}{2}.\label{e:A6} 
\end{eqnarray} 
\Eref{e:A6} can be simply generalized to the case where Alice-Bob share a $\PPR_{\xi}$ box, and Bob-Charlie share a $\PPR_{\xi'}$ box; it suffices to replace the term $(2\xi-1)^2$ by $(2\xi-1)(2\xi'-1)$.
In those cases where Bob is unsuccessful in swapping non-locality, that is when $b' = 1$, the final box held by Alice and Charlie, which we call the failure box and denote $P_\mathrm{f}$, is given by
\begin{equation}
P_\mathrm{f} = \frac{1}{P(b'=0)}\big(\mathds{1} - P(b'=0)P_\mathrm{S}(ac|xz)\big),
\end{equation}
which has CH value
\begin{equation}
\vec{\mathrm{CH}}\cdot \vec{P}_{\mathrm{S}}(ac|xz) = \case{1}{2}(\case{3}{2}-X_\mathrm{t}).
\end{equation}
There are two things to note. First this expression is independent of $X_\mathrm{b}$ and is therefore independent of the class of coupler. Second, it is always positive, and therefore the failure box is always a local box.
\section*{References}
\bibliographystyle{unsrt}
\bibliography{thesis}

\end{document}